\documentclass{aa}  

\usepackage{graphicx}
\usepackage{txfonts}
\begin{document}

   \title{The solar-wind angular-momentum flux observed during Solar Orbiter's first orbit}

  \author{Daniel Verscharen\inst{1,2} \and  
David Stansby\inst{1} \and
Adam J.~Finley\inst{3} \and
Christopher J.~Owen\inst{1} \and
Timothy Horbury\inst{4} \and
Milan Maksimovic\inst{5}\and
Marco Velli\inst{6} \and
Stuart D.~Bale\inst{4,7,8,9} \and
Philippe Louarn\inst{10} \and
Andrei Fedorov\inst{10} \and
Roberto Bruno\inst{11} \and
Stefano Livi\inst{12} \and
Yuri V.~Khotyaintsev\inst{13} \and
Antonio Vecchio\inst{5,14} \and
Gethyn R.~Lewis\inst{1} \and
Chandrasekhar Anekallu\inst{1} \and
Christopher W.~Kelly\inst{1} \and
Gillian Watson\inst{1} \and
Dhiren O.~Kataria\inst{1} \and
Helen O'Brien\inst{4} \and
Vincent Evans\inst{4} \and
Virginia Angelini\inst{4} \and
the Solar Orbiter SWA, MAG and RPW teams
}

   \institute{$^{1}$ Mullard Space Science Laboratory, University College London, Holmbury House, Holmbury St Mary, Dorking, RH6~6NT, UK     \email{d.verscharen@ucl.ac.uk} \\
$^{2}$  Space Science Center, University of New Hampshire, Durham NH 03824\\
$^{3}$    Department of Astrophysics-AIM, University of Paris-Saclay and University of Paris, CEA, CNRS, Gif-sur-Yvette Cedex 91191, France\\
$^{4}$    Department of Physics, The Blackett Laboratory, Imperial College London, London, SW7~2AZ, UK\\
$^{5}$ LESIA, Observatoire de Paris, Universit\'e PSL, CNRS, Sorbonne Universit\'e, Universit\'e de Paris, 5 Place Jules Janssen, 92195 Meudon, France\\
$^{6}$    Department of Earth, Planetary, and Space Sciences, University of California Los Angeles, Los Angeles, CA 90095, USA\\
$^{7}$ Physics Department, University of California, Berkeley, CA 94720, USA\\
$^{8}$    Space Sciences Laboratory, University of California, Berkeley, CA 94720, USA\\
$^{9}$    School of Physics and Astronomy, Queen Mary University of London, London, E1~4NS, UK\\
$^{10}$    Institut de Recherche en Astrophysique et Plan\'etologie, 31028 Toulouse Cedex 4, France\\
$^{11}$    INAF-Istituto di Astrofisica e Planetologia Spaziali, 00133 Roma, Italy \\
$^{12}$    Southwest Research Institute, San Antonio, TX 78238, USA \\
$^{13}$ Swedish Institute of Space Physics (IRF), Box 537, 751~21 Uppsala, Sweden \\
$^{14}$ Radboud Radio Lab, Department of Astrophysics/IMAPP-Radboud University, P.~O.~ Box 9010, 6500~GL Nijmegen, The Netherlands
             }

   \date{Received XXX; accepted XXX}

 
  \abstract
   {}
   {We present the first measurements of the solar-wind angular-momentum flux recorded by the Solar Orbiter spacecraft.  Our aim is the validation of these measurements to support future studies of the Sun's angular-momentum loss.}
   {We combine 60-minute averages of the proton bulk moments and the magnetic field measured by the Solar Wind Analyser (SWA) and the magnetometer (MAG) onboard Solar Orbiter. We calculate the angular-momentum flux per solid-angle element using data from the first orbit of the mission's cruise phase during 2020. We separate the contributions from protons and from magnetic stresses to the total angular-momentum flux. }
   {The angular-momentum flux varies significantly over time. The particle contribution typically dominates over the magnetic-field contribution during our measurement interval. The total angular-momentum flux shows the largest variation and is typically anti-correlated with the radial solar-wind speed. We identify a compression region, potentially associated with a co-rotating interaction region or a coronal mass ejection, that leads to a significant localised increase in the angular-momentum flux, yet without a significant increase in the angular momentum per unit mass. We repeat our analysis using the density estimate from the Radio and Plasma Waves (RPW) instrument. Using this independent method, we find a decrease in the peaks of positive angular-momentum flux but otherwise consistent results.}
   {Our results largely agree with previous measurements of the solar-wind angular-momentum flux in terms of amplitude, variability, and dependence on radial solar-wind bulk speed. Our analysis highlights the potential for future, more detailed, studies of the solar wind's angular momentum and its other large-scale properties with data from Solar Orbiter. We emphasise the need to study the radial evolution and latitudinal dependence of the angular-momentum flux in combination with data from Parker Solar Probe and assets at heliocentric distances of 1\,au and beyond.}


   \maketitle
%

\section{Introduction}

The solar corona expands into interplanetary space in the form of the solar wind \citep{parker58,neugebauer62,verscharen19}. In this process, the solar-wind plasma removes mass, momentum, energy, and angular momentum from the Sun. In the lower corona, the Sun's magnetic field forces the plasma into a quasi-rigid co-rotation with the photosphere, following the co-rotation of the field's photospheric footpoints. With increasing distance from the photosphere, the torque exerted by the coronal magnetic field on the plasma decreases  \citep{weber67}. 
In this way, the large-scale magnetic field mediates a smooth transition from co-rotation to quasi-radial expansion. Therefore, the solar wind is not ``flung''  from the photosphere on a ballistic trajectory, which would lead to a torque-free azimuthal velocity profile of the form 
\begin{equation}\label{Uphisimp}
U_{\phi}(r)\simeq \frac{R_{\odot}^2\Omega_{\odot}\sin\theta}{r}, 
\end{equation}
where $R_{\odot}$ is the solar radius, $\Omega_{\odot}$ is the Sun's angular rotation frequency, $\theta$ is the co-latitude, and $r$ is the heliocentric distance. Instead, the solar wind experiences a significant torque in its acceleration region at even larger distances from the photosphere, leading to greater azimuthal speeds than predicted by Eq.~(\ref{Uphisimp}). In turn, the torque applied to the Sun by the magnetic field in this process slows down the Sun's rotation on long time scales \citep{mestel68,reiners12}. Local measurements of the solar-wind angular-momentum flux provide an estimate for the global angular-momentum loss rate, which ultimately causes the rotation period of the Sun and, by extrapolation, of Sun-like stars on the main sequence to increase with age \citep{barnes03,gallet13,matt15,pantolmos17}.

Measurements of the solar-wind angular-momentum flux are particularly challenging from an instrumental point of view.  They require an accurate determination of the azimuthal component $U_{\phi}$ of the particle bulk velocity, which is typically more than one order of magnitude less than its radial component $U_r$. Nevertheless, early measurements of the solar wind already estimated its angular momentum \citep{hundhausen70,lazarus71,pizzo83,marsch84}. Modern space instrumentation provides us with higher spacecraft-pointing accuracy and thus a more accurate determination of $U_{\phi}$. Moreover, as the average ratio $U_{\phi}/U_r$ increases with decreasing heliocentric distance, the error in the quantification of the angular-momentum flux is generally smaller when measured at smaller distances from the Sun \citep{finley20,finley21}. 
In addition, co-rotating interaction regions have not yet formed \citep{richter86,allen20}, and interplanetary coronal mass ejections have not fully expanded to disturb nearby wind streams \citep{moestl20} at small distances from the Sun. Therefore, near-Sun measurements enable the sampling of more pristine and less processed solar wind.
Solar Orbiter and Parker Solar Probe share the advantages of both modern space instrumentation and an orbit that leads them close to the Sun \citep{fox16,mueller20}.
We present the first observations of the solar wind's angular-momentum flux observed by Solar Orbiter during the first orbit of its cruise phase in 2020.

\section{Data analysis}\label{sect_data}

We analyse a combined set of data from the Proton-Alpha Sensor (PAS) of Solar Orbiter's Solar Wind Analyser \citep[SWA,][]{owen20} suite and from Solar Orbiter's fluxgate magnetometer \citep[MAG,][]{horbury20}. Our dataset includes all time intervals from 2020-07-07 until 2020-10-27, for which both SWA/PAS and MAG data are available. We ignore all other intervals and those in which the quality flags for either dataset indicate poor data quality (we only include data with a quality flag of 3).  After this selection, we thus retain about 38.9\% of the total time interval in our dataset  (for time coverage, see also Figure~\ref{am_timeseries}).  During this time, Solar Orbiter recorded data at heliocentric distances between 0.591 and 0.989\,au. The time average of the spacecraft's heliocentric distance in our dataset is 0.851\,au. 

The calculation of the angular-momentum flux is very sensitive to knowledge uncertainties in the direction of the bulk velocity and in the direction of the magnetic field. The finite angular resolution of PAS introduces an uncertainty in the knowledge of the bulk-speed direction. The calibration accuracy of PAS is $\lesssim 1^{\circ}$. The pointing knowledge of the MAG sensor is largely determined by the uncertainty of the spacecraft's boom deployment angle. It introduces an angular uncertainty of the magnetic-field measurement of $\lesssim 1^{\circ}$.  Planned inter-instrument alignment reconstructions of the boom orientation will become feasible in the future once the instruments have recorded sufficient amounts of data for a large statistical analysis \citep{walsh20}.  For solar-wind speeds above $300\,\mathrm{km\,s}^{-1}$, the relative uncertainty of the SWA/PAS speed measurement due to counting statistics is less than 1\%.  The relative uncertainty of the density measurement is energy-dependent with a maximum of 20\%, which we apply as a conservative estimate to our entire dataset. At very low energies (corresponding to solar-wind speeds below $\sim 300\,\mathrm{km\,s}^{-1}$), the sensitivity decreases further so that additional correction factors are required, which have not been conclusively determined yet. The expected MAG offset is $\sim \pm 0.1\,\mathrm{nT}$, which we take as the absolute uncertainty of our individual magnetic-field component measurements.  The requirement of the absolute knowledge error of the spacecraft pointing is $\leq 3\,\mathrm{arcmin}$ \citep{garcia21}, which is mostly driven by the instrument requirements of the remote-sensing suite and a minor contributor to the pointing knowledge uncertainties for our study.

We base our analysis on the PAS normal-mode ground moments integrated from the proton part (core and beam) of the full three-dimensional measured ion distributions and the MAG normal-mode vectors of the magnetic field. In normal mode,  the PAS ground moments are available every 4\,seconds, while the MAG vectors are available with a cadence of 8\,vectors per second.  We merge and average both data products over intervals of length 60~min to reduce natural fluctuations in the data due to, for example, waves and turbulence, as we are interested in the angular-momentum flux of the bulk solar wind. Our selection and averaging procedure leaves us with 1036 individual data points.

The angular momentum is contained in the mechanical flux of the solar-wind particles. In addition, there is transport of angular momentum due to magnetic-field stresses. Ignoring anisotropies in the particle distributions and contributions from particles other than protons\footnote{PAS has the capability to determine the moments of the solar-wind $\alpha$-particle component as well. However, this dataset requires further calibration, so that we neglect $\alpha$-particles at this point.}, the total angular-momentum flux is thus the sum of the proton and magnetic-field terms \citep[see][]{marsch84}. We define the total angular-momentum flux per solid-angle element in spherical heliocentric coordinates as
\begin{equation}\label{angmomtot}
\mathcal F_{\mathrm{tot}}=\mathcal F_{\mathrm p}+\mathcal F_B,
\end{equation}
where
\begin{equation}\label{angmomp}
\mathcal F_{\mathrm p}=r^3\rho U_rU_{\phi}
\end{equation}
is the proton contribution to the angular-momentum flux,
\begin{equation}\label{angmomB}
\mathcal F_B=-r^3\frac{B_rB_{\phi}}{4\pi}
\end{equation}
is the contribution from magnetic stresses (in cgs units) to the angular-momentum flux, $r$ is the heliocentric distance, $\rho$ is the proton mass density, $\vec U$ is the proton bulk velocity, and $\vec B$ is the magnetic field. The subscript $r$ indicates the radial vector component, and the subscript $\phi$ indicates the azimuthal component in spherical heliocentric coordinates. We use our 60-minute averages of the measured solar-wind parameters as the input for our calculations of $\mathcal F_{\mathrm p}$ and $\mathcal F_B$ according to Equations~(\ref{angmomp}) and (\ref{angmomB}). We recognise that our analysis represents only a first validation of methods to study angular-momentum flux with Solar Orbiter. Therefore, it ignores other contributions to the total angular-momentum flux (see Section~\ref{sect_disc}), which must be investigated in future detailed studies.

\section{Results}

\subsection{Timeseries and overview}

\begin{figure*}
 \centering
\includegraphics[width=\textwidth]{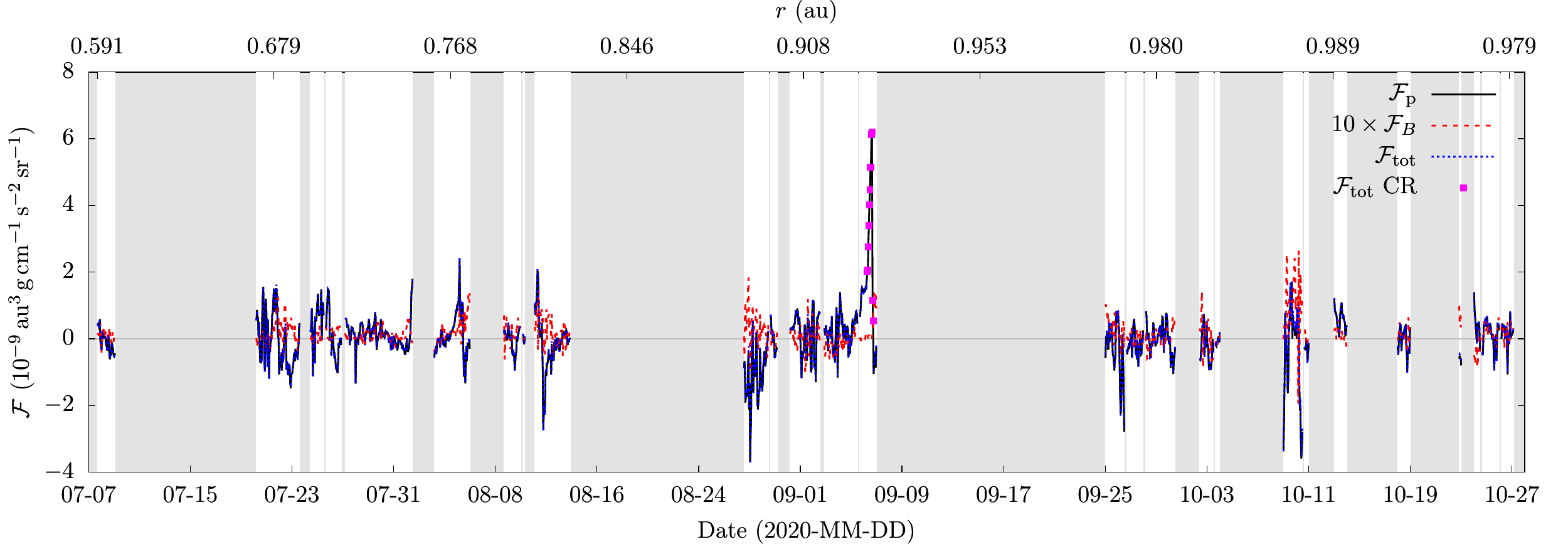}
\caption{Timeseries of the angular-momentum flux per solid-angle element measured during Solar Orbiter's first orbit of the cruise phase. The diagram shows the proton contribution $\mathcal F_{\mathrm p}$, the magnetic-field contribution $\mathcal F_B$ and the sum of both, $\mathcal F_{\mathrm{tot}}$. In order to increase the visibility, we multiply $\mathcal F_B$ with a factor 10. The magenta squares represent the measurements of $\mathcal F_{\mathrm{tot}}$ during the time of the compression region (CR) on 2020-09-06. The axis at the top indicates the heliocentric distance of the spacecraft at the time of the measurement. The scale of this axis is not linear. The grey-shaded areas indicate times for which our merged SWA/PAS-MAG data product is unavailable or the data flags for either dataset indicate poor data quality.}
   \label{am_timeseries}
 \end{figure*}

Figure~\ref{am_timeseries} shows the timeseries of $\mathcal F_{\mathrm p}$, $\mathcal F_B$, and $\mathcal F_{\mathrm{tot}}$ over the analysed data interval. Gaps in this plot represent data gaps or those intervals that we exclude according to our selection criteria. Figure~\ref{am_timeseries} illustrates the natural variation of the angular-momentum flux over time, which is typically greater than its mean magnitude. Across all data points, the mean value of $\mathcal F_{\mathrm p}$ is  $2.29\times 10^{-11}\,\mathrm{au}^3\,\mathrm{g\,cm}^{-1}\,\mathrm s^{-2}\,\mathrm{sr}^{-1}$, while the mean value of $\mathcal F_B$ is $1.72\times 10^{-11}\,\mathrm{au}^3\,\mathrm{g\,cm}^{-1}\,\mathrm s^{-2}\,\mathrm{sr}^{-1}$. In general, $\mathcal F_{\mathrm p}$ exhibits significantly more variation than $\mathcal F_B$. As expected for a conserved quantity, $\mathcal F_{\mathrm{tot}}$ shows no secular dependence on heliocentric distance. 

On 2020-09-06, we record a time interval of significantly increased $\mathcal F_{\mathrm p}$ and $\mathcal F_{\mathrm {tot}}$. Upon closer inspection, this interval corresponds to a time of increased $\rho$, probably associated with the compression in front of a co-rotating interaction region or a coronal mass ejection seen as a flux rope in the magnetic field. Due to this enhancement in $\rho$, the associated plasma carries more angular-momentum flux than the solar wind before or after the compression region. The compression region does not exhibit enhancements in $U_r$ or $U_{\phi}$ (not shown here), suggesting that the compression region facilitates a similar angular-momentum flux per unit mass as the solar wind. Figure~\ref{am_massflux} supports this suggestion as discussed below. We highlight the measurements taken during the time interval associated with this compression region in our figures as magenta squares.

\subsection{Variability of the angular-momentum flux}

\begin{figure}
 \centering
\includegraphics[width=\columnwidth]{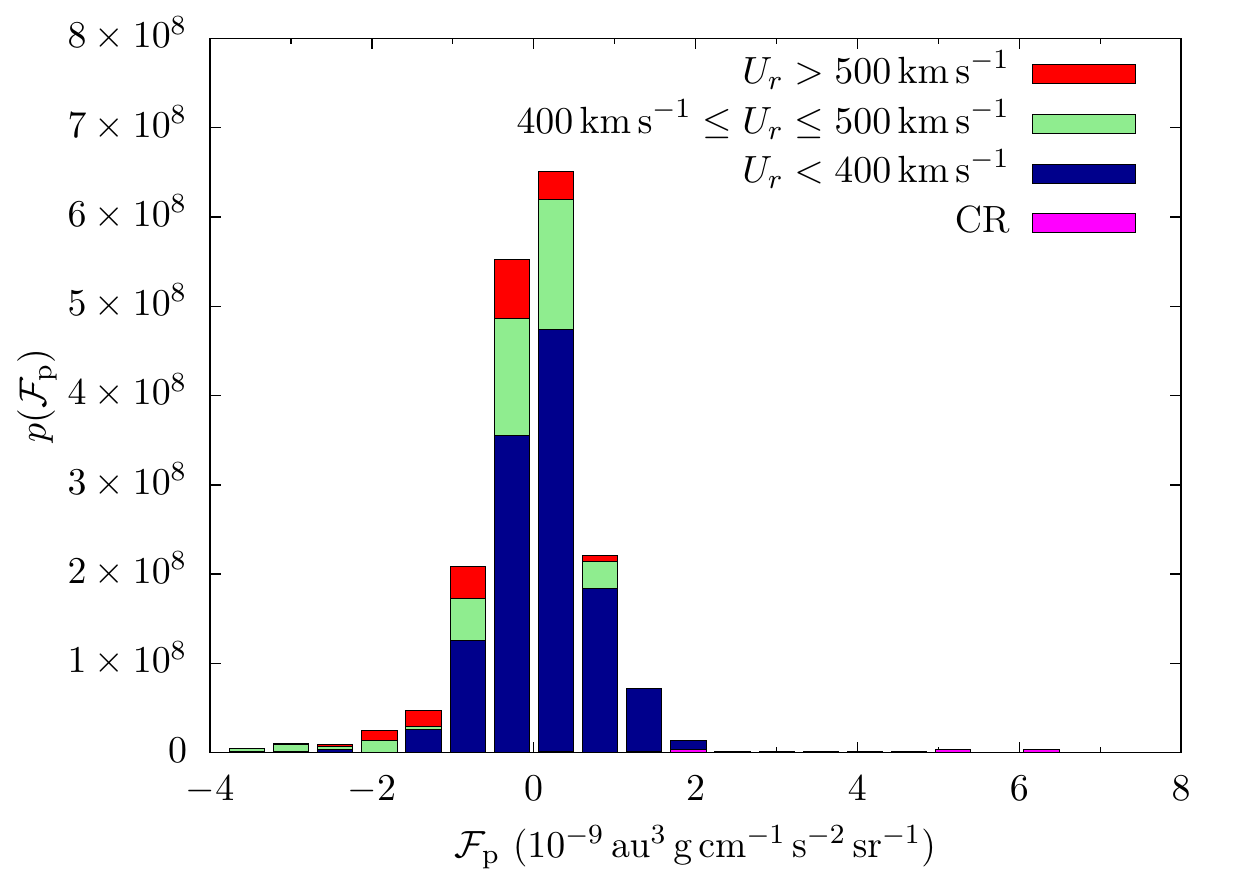}
\caption{Histogram of the  proton contribution $\mathcal F_{\mathrm p}$ to the solar wind's angular-momentum flux per solid-angle element. The vertical axis indicates the probability density of $\mathcal F_{\mathrm p}$. We stack the histograms for fast ($U_r>500\,\mathrm{km\,s}^{-1}$), intermediate ($400\,\mathrm{km\,s}^{-1}\le U_r\le 500\,\mathrm{km\,s}^{-1}$), and slow ($U_r<400\,\mathrm{km\,s}^{-1}$) wind. The magenta areas represent the measurements of $\mathcal F_{\mathrm{p}}$ during the time of the compression region (CR) on 2020-09-06.}
   \label{hist_angmom_part}
 \end{figure}

\begin{figure}
 \centering
\includegraphics[width=\columnwidth]{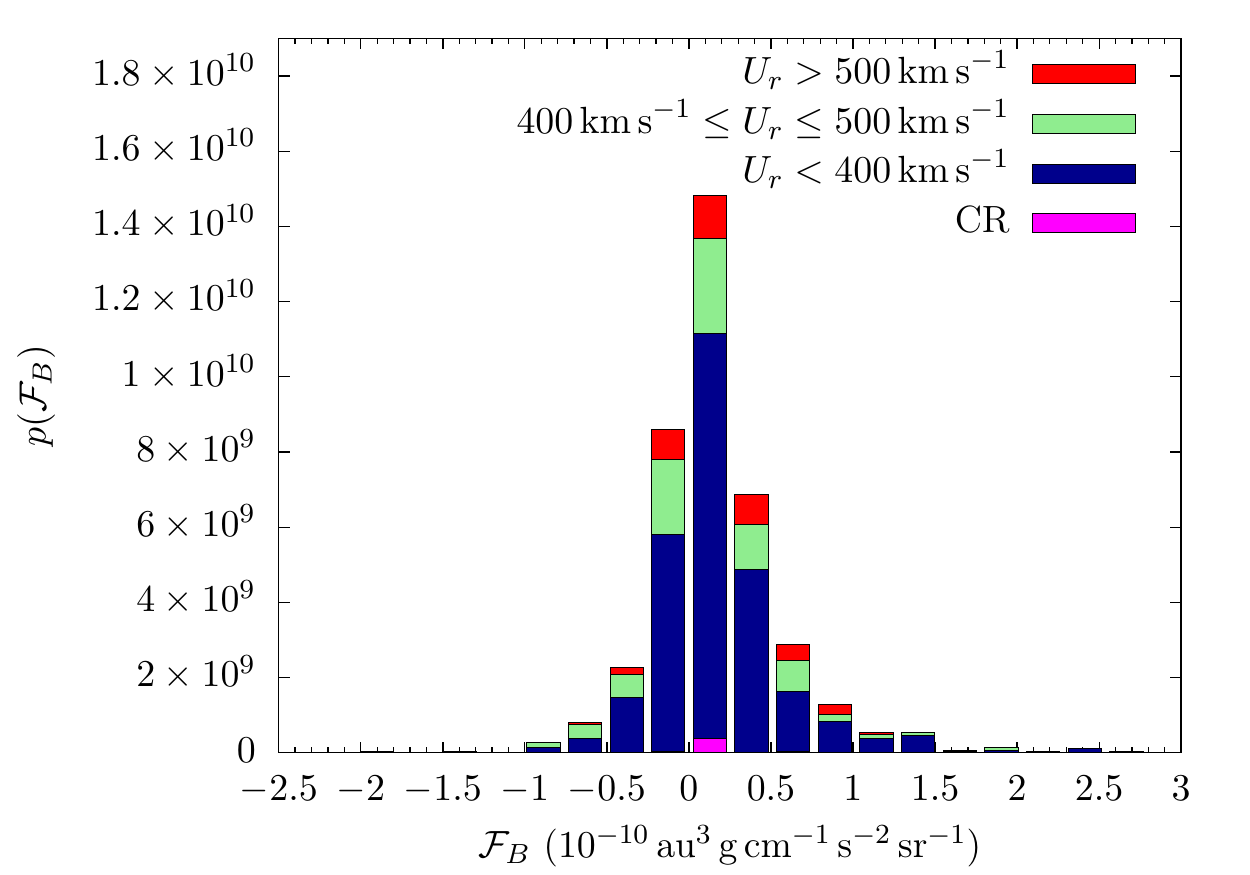}
\caption{Same as Figure~\ref{hist_angmom_part}, but for the magnetic-field contribution $\mathcal F_{B}$ to the solar wind's angular-momentum flux per solid-angle element. }
   \label{hist_angmom_field}
 \end{figure}

In Figures~\ref{hist_angmom_part} and \ref{hist_angmom_field}, we show histograms of the measured values of $\mathcal F_{\mathrm p}$ and $\mathcal F_{B}$ in our dataset in terms of their probability distributions. We colour-code the contributions from fast, intermediate, and slow solar wind in our histograms. Due to the dominance of $\mathcal F_{\mathrm p}$, the histogram of the probability distribution for $\mathcal F_{\mathrm{tot}}$ (not shown) is almost identical with the histogram for $\mathcal F_{\mathrm p}$. Both histograms reflect the large variation in the angular-momentum flux and the range of observed values.  According to Figure~\ref{hist_angmom_part}, the majority distribution is centred around negative values of $\mathcal F_{\mathrm p}$, while the outliers at large $\mathcal F_{\mathrm p}$ shift the mean of our measurements to $\mathcal F_{\mathrm p}>0$. These outliers are mostly associated with the compression region on 2020-09-06.

%

For a more quantitative statistical analysis, we list the mean values, the standard deviations, as well as the maximum and minimum values of $\mathcal F_{\mathrm p}$, $\mathcal F_B$, and $\mathcal F_{\mathrm{tot}}$ in Table~\ref{table_stats}. In addition, Table~\ref{table_stats} provides these statistical markers separated by time intervals with $U_r<400\,\mathrm{km\,s}^{-1}$ and $U_r>500\,\mathrm{km\,s}^{-1}$ as a means to distinguish characteristic slow wind from fast wind, respectively. We find that $\mathcal F_{\mathrm p}>0$ and $\mathcal F_{\mathrm {tot}}>0$ in the observed slow-wind intervals ($U_r<400\,\mathrm{km\,s}^{-1}$), while  $\mathcal F_{\mathrm p}<0$ and $\mathcal F_{\mathrm {tot}}<0$ in the observed fast-wind intervals ($U_r>500\,\mathrm{km\,s}^{-1}$).
A similar behaviour has been seen in Parker Solar Probe measurements \citep{finley21}.
In the separated data subsets for slow wind and fast wind, the mean values of $|\mathcal F_{\mathrm p}|$ are greater than the mean value of $|\mathcal F_{\mathrm p}|$ in our full dataset of all measurements.  
The means of $\mathcal F_B$ are positive, independent of our categorisation by wind speed,
and the mean value of $\mathcal F_B$ in our slow-wind and fast-wind intervals is less than the mean value of $\mathcal F_B$ in all measurement intervals.  We note, however, that the mean value of $\mathcal F_B$ is very small for the separated slow-wind and fast-wind intervals, so that this result is potentially not significant (see also Figure~\ref{am_speed_field}).

\begin{table*}
 \caption{Statistical properties of $\mathcal F_{\mathrm p}$, $\mathcal F_B$, and $\mathcal F_{\mathrm{tot}}$ in our full dataset and split by radial solar-wind speed.}
 \label{table_stats}
 \centering
 \begin{tabular}{lccc}
 \hline \hline
 & mean & min & max \\
  &$(10^{-11}\,\mathrm{au}^3\,\mathrm g\,\mathrm{cm}^{-1}\,\mathrm s^{-2}\,\mathrm{sr}^{-1})$ & $(10^{-9}\,\mathrm{au}^3\,\mathrm g\,\mathrm{cm}^{-1}\,\mathrm s^{-2}\,\mathrm{sr}^{-1})$&$(10^{-9}\,\mathrm{au}^3\,\mathrm g\,\mathrm{cm}^{-1}\,\mathrm s^{-2}\,\mathrm{sr}^{-1})$ \\
 \hline
 all speeds & & & \\
 $\mathcal F_{\mathrm{p}}$ & $2.29\pm 82.57$ & $-3.66$ & $6.19$  \\
$  \mathcal F_{B}$ & $1.72\pm4.16$ & $-0.194$ & $0.264$ \\
$ \mathcal F_{\mathrm{tot}}$ & $4.01\pm82.59$ & $-3.75$ & $6.19$ \\
 \hline
 $U_r<400\,\mathrm{km\,s}^{-1}$ & & & \\
 $\mathcal F_{\mathrm{p}}$ & $15.5\pm 81.2$ & $-3.37$ & $6.19$  \\
$  \mathcal F_{B}$ & $0.923\pm5.560$ & $-0.230$ & $0.254$ \\
$ \mathcal F_{\mathrm{tot}}$ & $16.4\pm81.0$ & $-3.37$ & $6.19$ \\
 \hline
  $U_r>500\,\mathrm{km\,s}^{-1}$ &  & & \\
 $\mathcal F_{\mathrm{p}}$ & $-73.4\pm 119.8$ & $-5.48$ & $0.834$  \\
$  \mathcal F_{B}$ & $0.593\pm5.474$ & $-0.174$ & $0.101$ \\
$ \mathcal F_{\mathrm{tot}}$ & $-72.8\pm121.8$ & $-5.49$ & $0.868$ \\
 \hline
 \end{tabular}
 \end{table*}

\subsection{Speed dependence of the angular-momentum flux}\label{sect_speed}

\begin{figure}
 \centering
\includegraphics[width=\columnwidth]{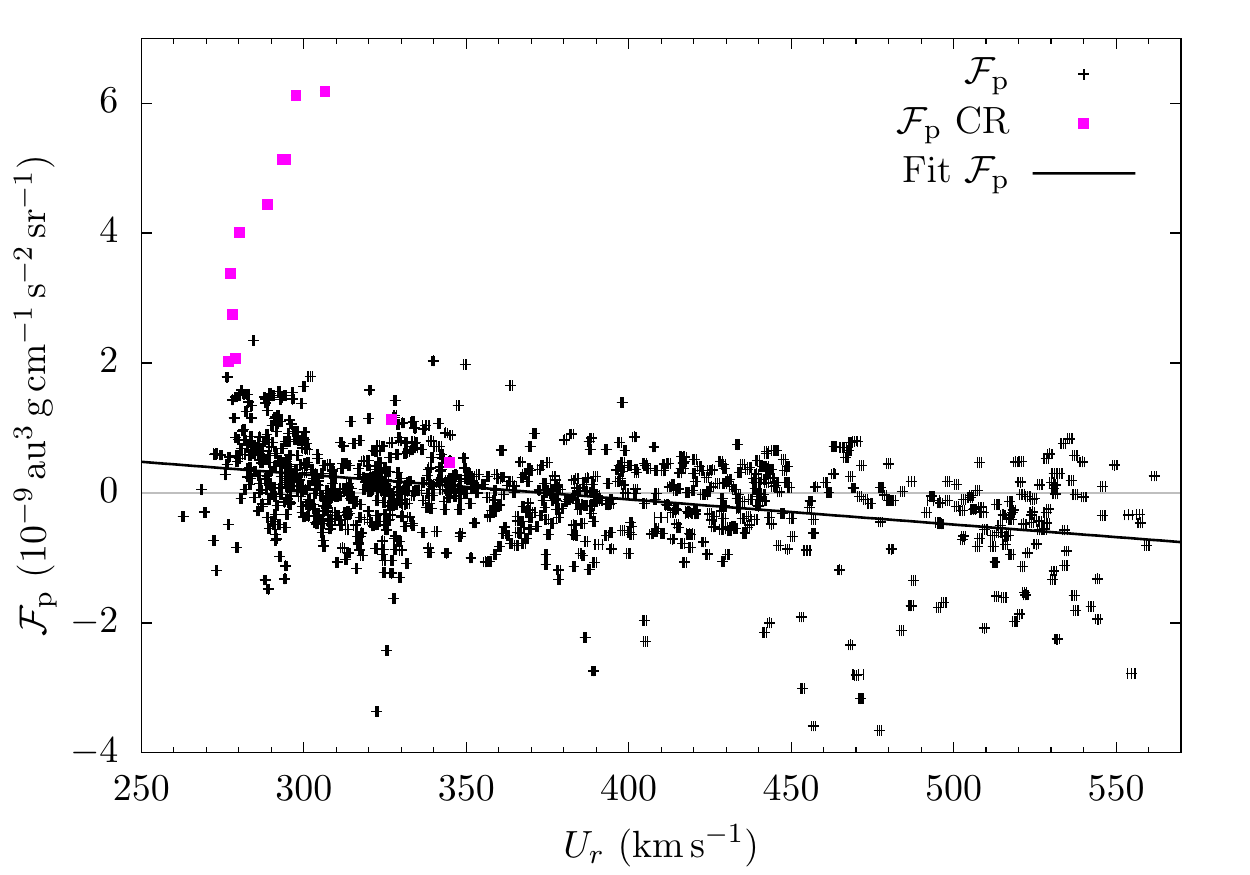}
\caption{Dependence of the proton contribution $\mathcal F_{\mathrm p}$ to the angular-momentum flux on the radial proton bulk speed $U_r$. We show the measurements of $\mathcal F_{\mathrm p}$ as points and overplot the linear fit to Eq.~(\ref{speedeq}) as a black line.  All measurement points have horizontal and vertical error bars, which are small due to the time-averaging. The fit parameters are given in Table~\ref{table_fit}.  The magenta squares represent the measurements of $\mathcal F_{\mathrm{p}}$ during the time of the compression region (CR) on 2020-09-06.}
   \label{am_speed}
 \end{figure}
  \begin{figure}
 \centering
\includegraphics[width=\columnwidth]{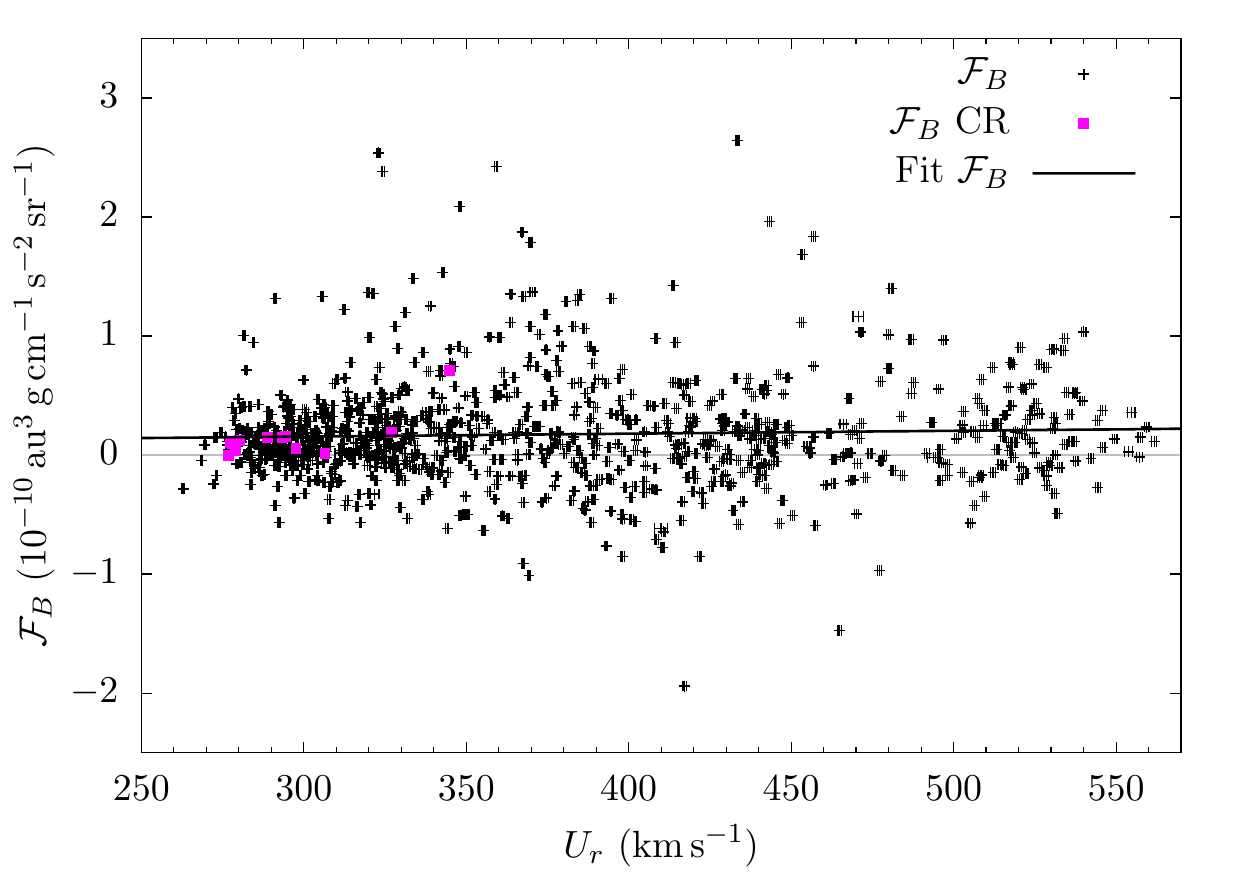}
\caption{Dependence of the magnetic-field contribution $\mathcal F_{B}$ to the angular-momentum flux on the radial proton bulk speed $U_r$. We show the measurements of $\mathcal F_{B}$ as points and overplot the linear fit to Eq.~(\ref{speedeq}) as a black line.  All measurement points have horizontal and vertical error bars, which are small due to the time-averaging. The fit parameters are given in Table~\ref{table_fit}.  The magenta squares represent the measurements of $\mathcal F_{B}$ during the time of the compression region (CR) on 2020-09-06. }
   \label{am_speed_field}
 \end{figure}
Figure~\ref{am_speed} shows our measurements of $\mathcal F_{\mathrm p}$ as a function of $U_r$. This visualisation confirms the significant $U_r$-dependence of the angular-momentum flux, which has been noted in previous studies and is discussed further in Section~\ref{sect_disc}. The particle contribution to the angular-momentum flux shows a stronger relative variation in slow wind compared to fast wind. We also observe a trend towards negative values of $\mathcal F_{\mathrm p}$ (and thus $\mathcal F_{\mathrm{tot}}$) at larger $U_r$. Figure~\ref{am_speed_field} shows the same but for $\mathcal F_{B}$. The magnetic-field contribution does not follow the same clear trend in its $U_r$-dependence as $\mathcal F_{\mathrm p}$. Figures \ref{am_speed} and \ref{am_speed_field} include horizontal and vertical error bars to represent $\Delta U_r$, $\Delta \mathcal F_{\mathrm p}$, and $\Delta \mathcal F_B$. The error bars follow from error propagation of the individual measurement uncertainties quoted in Section~\ref{sect_data}, which corresponds to the application of the standard error of the mean based on the individual uncertainties. We find that, in our hourly averages, the relative errors $\Delta U_r$, $\Delta \mathcal F_{\mathrm p}$, and $\Delta \mathcal F_B$ are negligible.

A comparison of the magenta points in Figures~\ref{am_speed} and \ref{am_speed_field} reveals that $\mathcal F_{B}$ is not enhanced in the compression region compared to the time intervals outside the compression region. We note that the unit on the vertical axis in Figure~\ref{am_speed_field} is one order of magnitude smaller than the unit on the vertical axis in Figure~\ref{am_speed}. This reflects again that $\mathcal F_{\mathrm p}$ is on average the dominant contribution to $\mathcal F_{\mathrm{tot}}$. In order to quantify the $U_r$-dependence, we apply least-squares Marquardt--Levenberg fits to our measurements according to the linear equation
\begin{equation}\label{speedeq}
\mathcal F=aU_r+b
\end{equation}
with the fit parameters $a$ and $b$, where $\mathcal F$ represents either  $\mathcal F_{\mathrm p}$, $\mathcal F_{B}$, or $\mathcal F_{\mathrm{tot}}$. 
We show the resulting fit parameters including their errors in Table~\ref{table_fit}.

  \begin{table}
 \caption{Results of our linear fits to the $U_r$-dependence of $\mathcal F_{\mathrm p}$, $\mathcal F_{B}$, and $\mathcal F_{\mathrm{tot}}$ according to Eq.~(\ref{speedeq}).}
 \label{table_fit}
 \centering
 \begin{tabular}{ccc}
 \hline \hline
  & $a$ & $b$ \\ 
 &  $(10^{-17}\,\mathrm{au}^3\,\mathrm g\,\mathrm{cm}^{-2}\,\mathrm s^{-1}\,\mathrm{sr}^{-1})$ & $(10^{-9}\,\mathrm{au}^3\,\mathrm g\,\mathrm{cm}^{-1}\,\mathrm s^{-2}\,\mathrm{sr}^{-1})$\\
\hline
 $\mathcal F_{\mathrm{p}}$ & $-3.86\pm0.32 $ & $1.45\pm 0.12 $  \\
$  \mathcal F_{B}$ & $0.024\pm0.017$ & $0.0082\pm 0.0065$  \\
$ \mathcal F_{\mathrm{tot}}$ & $-3.84\pm0.32$ & $1.45\pm 0.12$  \\
 \hline
 \end{tabular}
 \end{table}

\subsection{Mass-flux dependence of the angular-momentum flux}

\begin{figure}
 \centering
\includegraphics[width=\columnwidth]{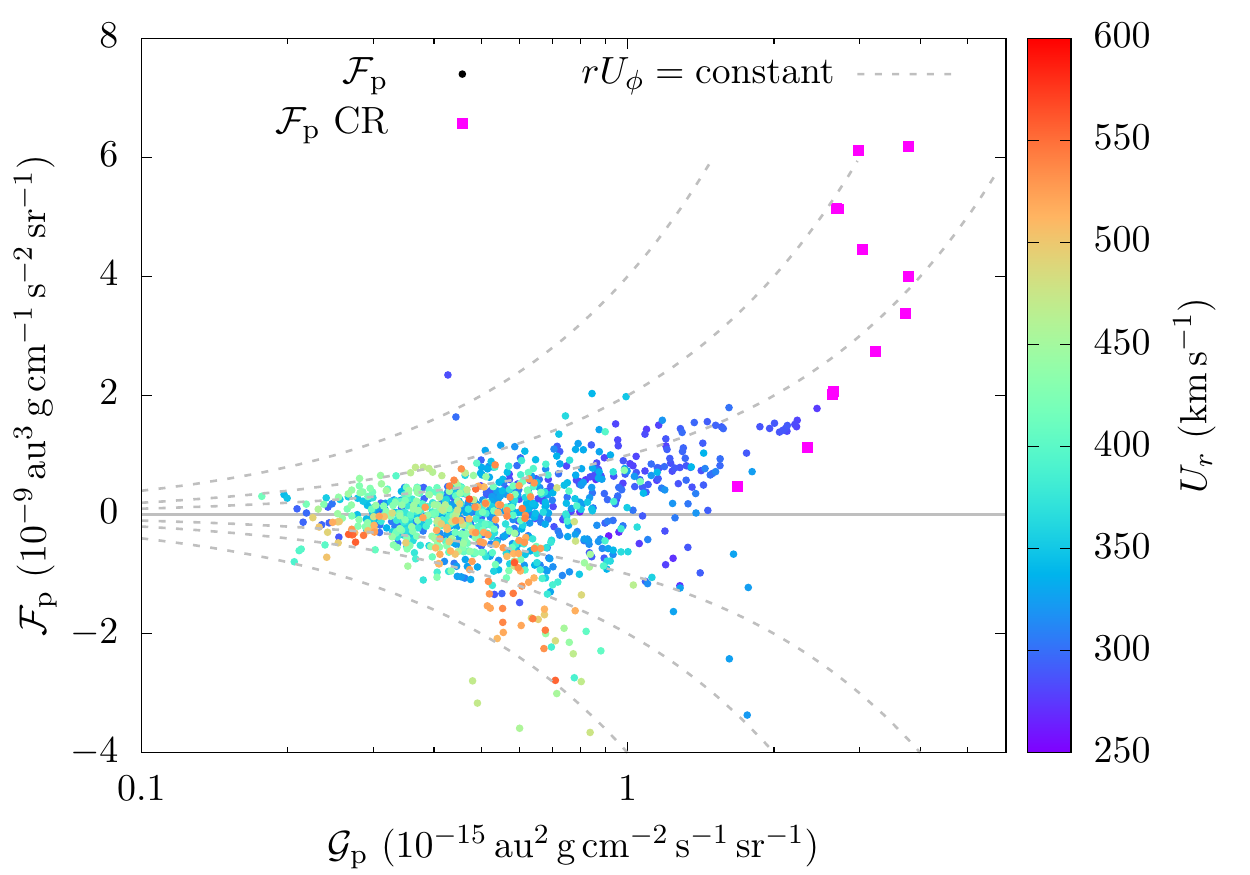}
\caption{Dependence of the proton contribution $\mathcal F_{\mathrm p}$ to the angular-momentum flux on the radial proton mass flux $\mathcal G_{\mathrm p}$. The point colour represents the value of $U_{r}$ for each measurement. The magenta squares represent the measurements of $\mathcal F_{\mathrm{p}}$ during the time of the compression region (CR) on 2020-09-06. The grey dashed lines indicate contours of constant $rU_{\phi}$.}
   \label{am_massflux}
 \end{figure}

We define the proton mass flux per solid-angle element as
\begin{equation}
\mathcal G_{\mathrm p}=r^2\rho U_r.
\end{equation}
As $\mathcal F_{\mathrm p}$ is carried by the proton flow, we combine $\mathcal F_{\mathrm p}$ and $\mathcal G_{\mathrm p}$ in Figure~\ref{am_massflux} and analyse their dependence. We colour-code each measurement point with its associated value of $U_{r}$, which allows us to link Figure~\ref{am_massflux} with Figure~\ref{am_speed}. 
All time intervals with $\mathcal G_{\mathrm p}\gtrsim 2\times 10^{-15}\,\mathrm{au}^2\,\mathrm{g\,cm}^{-2}\,\mathrm s^{-1}\,\mathrm{sr}^{-1}$ in our dataset exhibit $\mathcal F_{\mathrm p}>0$. The majority of these points correspond to the compression region on 2020-09-06. We highlight them as magenta squares in Figure~\ref{am_massflux}, illustrating that the mass-flux dependence is a useful method to separate transient and atypical plasma intervals from the regular background solar wind \citep[see also][]{stansby19}.

The value of $rU_{\phi}$ is a measure of the local specific angular momentum per proton. Since $\mathcal F_{\mathrm p}=rU_{\phi}\mathcal G_{\mathrm p}$, isocontours of constant $rU_{\phi}$ correspond to bent curves when using a logarithmic $\mathcal G_{\mathrm p}$-axis in Figure~\ref{am_massflux}. We show these isocontours as grey dashed curves for a range of $rU_{\phi}$ values from $\pm 10$ to $\pm40 \,\mathrm{au\,km\,s}^{-1}$. 
At the low-$\mathcal G_{\mathrm p}$ end,  the distribution scatters almost symmetrically  around a value of $\mathcal F_{\mathrm p}=0$. The envelope of the distribution in this $\mathcal G_{\mathrm p}$ range follows isocontours of constant $\pm \left|rU_{\phi}\right|$, meaning that the scatter in $\mathcal F_{\mathrm p}$  is well defined by a fixed range of constant magnitudes of the angular momentum per proton. 
At intermediate $\mathcal G_{\mathrm p}$, the symmetry around $\mathcal F_{\mathrm p}=0$ breaks, and more data points occur at $\mathcal F_{\mathrm p}>0$. At $\mathcal F_{\mathrm p}>0$, the envelope of the data distribution in terms of  $rU_{\phi}$  decreases as $\mathcal G_{\mathrm p}$ increases in the intermediate-$\mathcal G_{\mathrm p}$ range.
The points representing the compression region largely lie on the same isocontours of $rU_{\phi}$  as the bulk of the slow solar wind. This behaviour suggests that the increase in $\mathcal F_{\mathrm p}$ associated with the compression region is mostly due to an increase in $\rho$ rather than to an increase in the specific angular momentum per proton compared to the regular slow wind.


\subsection{Spectral analysis of angular-momentum flux}\label{sect_fourier}

\begin{figure}
 \centering
\includegraphics[width=\columnwidth]{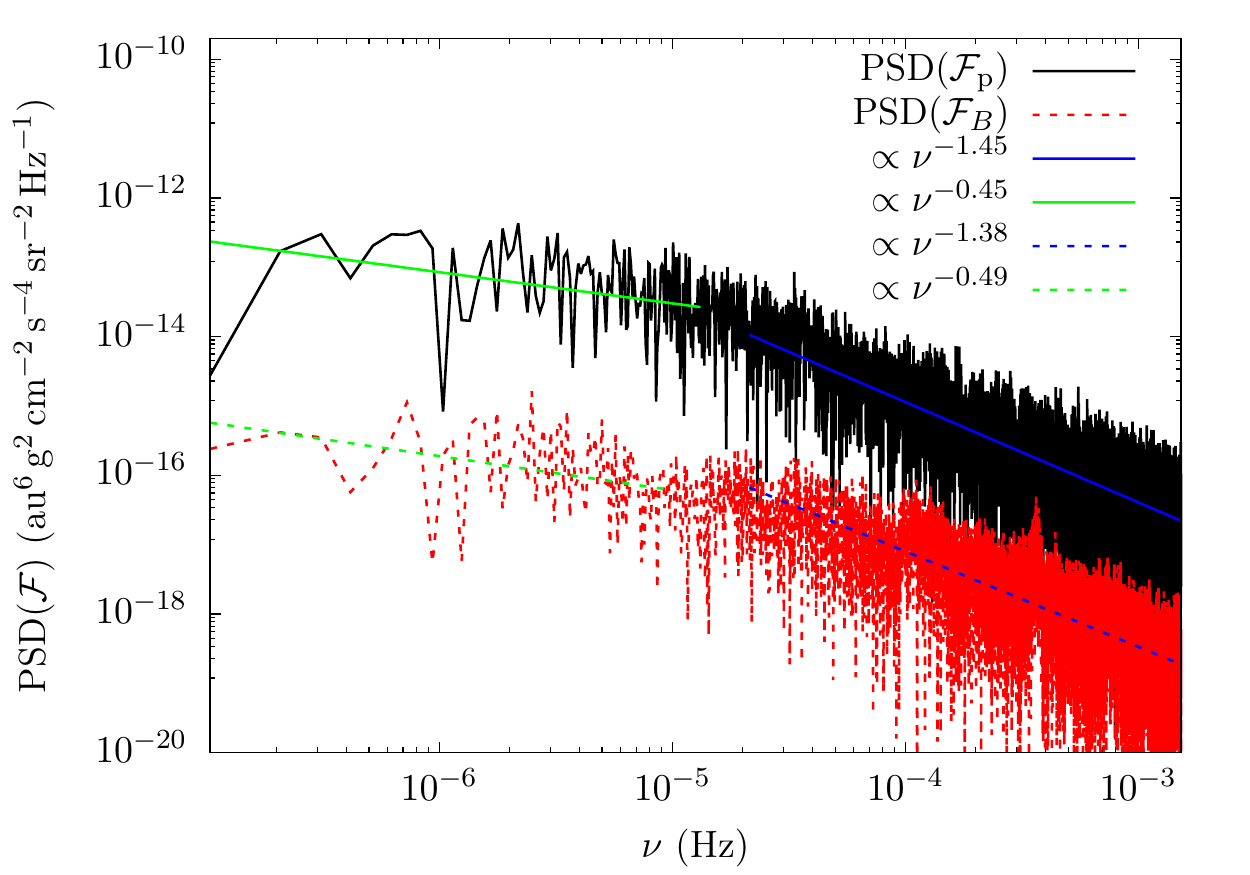}
\caption{Power spectral densities (PSDs) of $\mathcal F_{\mathrm p}$ and $\mathcal F_{B}$. The straight lines show power-law fits of the partial spectra of  $\mathcal F_{\mathrm p}$ and $\mathcal F_{B}$.}
   \label{fourier}
 \end{figure}

Solar Orbiter's high time resolution in the measurement of both the particles and the magnetic field enables the analysis of the power spectrum of the variable angular-momentum flux over a wide range of frequencies. In order to demonstrate the suitability of Solar Orbiter data for future studies of this type, we present such a spectral analysis of the angular-momentum flux based on our dataset.
 We apply a non-uniform fast Fourier transform \citep{barnett19,barnett21} to the timeseries data of $\mathcal F_{\mathrm p}$ and $\mathcal F_B$ throughout our entire dataset. For this calculation, we use 1-minute averages of the data instead of the 60-minute averages used in our study otherwise. This choice allows us to explore the variability of $\mathcal F_{\mathrm p}$ and $\mathcal F_B$ over a wider range of frequencies. 
Figure~\ref{fourier} shows the resulting power spectral densities (PSDs) of $\mathcal F_{\mathrm p}$ and $\mathcal F_B$ as functions of frequency $\nu$.  
 
 We observe a spectral break in the PSDs of both contributions to the angular-momentum flux near a frequency of $\nu\approx 10^{-5}\,\mathrm{Hz}$. We perform separate power-law fits to PSD($\mathcal F_{\mathrm p}$) and PSD($\mathcal F_B$) on both sides of their spectral breakpoints. We find PSD($\mathcal F_{\mathrm p}$)$\propto \nu^{-0.45\pm 0.12}$ in the low-$\nu$ regime and PSD($\mathcal F_{\mathrm p}$)$\propto \nu^{-1.448\pm 0.012}$ in the high-$\nu$ regime. Likewise, PSD($\mathcal F_{B}$)$\propto \nu^{-0.49\pm 0.14}$ in the low-$\nu$ regime and PSD($\mathcal F_{B}$)$\propto \nu^{-1.377\pm 0.012}$ in the high-$\nu$ regime. We overplot the fit results in Figure~\ref{fourier}.

\section{Discussion}\label{sect_disc}

Our measurements of $|\mathcal F_{\mathrm{p}}|$ are largely in agreement with previous measurements in the solar wind, which reveal values of about $4.4\dots 5.9\times10^{-11}\,\mathrm{au}^3\,\mathrm{g\,cm}^{-1}\,\mathrm s^{-2}\,\mathrm{sr}^{-1}$ \citep{pizzo83},  $4.4\dots 11.2\times10^{-11}\,\mathrm{au}^3\,\mathrm{g\,cm}^{-1}\,\mathrm s^{-2}\,\mathrm{sr}^{-1}$ \citep{finley20}, and  $1.09\dots 2.43\times10^{-10}\,\mathrm{au}^3\,\mathrm{g\,cm}^{-1}\,\mathrm s^{-2}\,\mathrm{sr}^{-1}$ \citep{liu21} as averages over their whole datasets. 

As shown in Section~\ref{sect_speed}, we find a general trend of decreasing $\mathcal F_{\mathrm{p}}$ with increasing $U_r$, consistent with previous findings in data from Wind and Parker Solar Probe \citep{finley19,finley20,finley21} and with estimates based on the observation of the Earth's magnetotail deflection \citep{nemecek20}. The observation of this trend in near-Sun data suggests that the trend is generated close to the Sun and not by local deflections at large distances. 
On average, $\mathcal F_{\mathrm{p}}>0$ for $U_r\lesssim 376\,\mathrm{km\,s}^{-1}$ and $\mathcal F_{\mathrm{p}}<0$ for $U_r\gtrsim 376\,\mathrm{km\,s}^{-1}$ according to our fit result applied to the data shown in Figure~\ref{am_speed}. The magnetic-field contribution $\mathcal F_B$ shows only a small dependence on $U_r$, which is consistent with recent measurements from Parker Solar Probe \citep{liu21}.

Averaged over very long timescales, we expect both the particle contribution and the magnetic-field contribution to the angular-momentum flux to be positive due to the Sun's sense of rotation and the direction of the average \citet{parker58} field. In addition, we anticipate an overall loss of the Sun's angular momentum based on observed trends in the rotation periods of Sun-like stars, which decrease with age along the main sequence. This finding is generally interpreted as a consequence of magnetised stellar winds \citep{lorenzo19,nascimento20}.
 In the Sun's specific case, a positive net $\mathcal F_{\mathrm{tot}}$ corresponds to a net loss of the Sun's rotational angular momentum. 
 
 Moreover, we expect the average relative contribution of $\mathcal F_B$ to the total angular-momentum flux to decrease slightly with increasing heliocentric distance as magnetic stress is converted into particle angular momentum \citep{weber67}, although \citet{marsch84} do not find clear evidence for this behaviour within 1\,au.
Previous measurements of $\mathcal F_B$ reveal values of about $4.1\dots 4.7\times10^{-11}\,\mathrm{au}^3\,\mathrm{g\,cm}^{-1}\,\mathrm s^{-2}\,\mathrm{sr}^{-1}$ \citep{pizzo83},  $3.6\dots 4.7\times10^{-11}\,\mathrm{au}^3\,\mathrm{g\,cm}^{-1}\,\mathrm s^{-2}\,\mathrm{sr}^{-1}$ \citep{finley20}, and  $2.1\dots 3.8\times10^{-11}\,\mathrm{au}^3\,\mathrm{g\,cm}^{-1}\,\mathrm s^{-2}\,\mathrm{sr}^{-1}$ \citep{liu21} as averages over their whole datasets at small heliocentric distances. These values are indeed greater than our mean values measured at Solar Orbiter's distances from the Sun, consistent with an average transfer of angular momentum from the field to the particles taking place between 0.3 and 0.7\,au. Averaged over two solar cycles, \citet{finley19} find $\mathcal F_B\approx 3.58\times 10^{-11}\,\mathrm{au}^3\,\mathrm{g\,cm}^{-1}\,\mathrm s^{-2}\,\mathrm{sr}^{-1}$ in data from the Wind spacecraft at 1\,au. Comparing to the averages at small heliocentric distances, this 1-au long-term average is also consistent with an overall decreasing trend of $\mathcal F_B$ with increasing $r$. We note, however, that our average $\mathcal F_B$ at intermediate distances is even smaller than the 1-au long-term average from Wind. This discrepancy is probably the result of the limited statistics in our dataset and the specific solar wind streams encountered by Solar Orbiter during its first orbit at this particular phase of the solar cycle.
Radial alignments between Solar Orbiter, Parker Solar Probe, and potentially other assets with suitable pointing accuracy and instrumentation will help us to understand the partition of the contributions to the angular-momentum flux over longer time intervals in the future. In general, we expect a significant dependence of the angular-momentum flux on the Sun's 11-year activity cycle based on changes to the global magnetic field strength and changing solar-wind source regions \citep{finley18}.

Notwithstanding these expectations regarding large-scale behaviour, our measurements reveal the large variation of the angular-momentum flux over many timescales from a few hours to months.  Taking our full measurement time interval into account, the relative variations in $\mathcal F_{\mathrm p}$ are greater than the relative variations in $\mathcal F_B$. When separating fast and slow wind, however, the relative variations in $\mathcal F_B$ are greater than those in $\mathcal F_{\mathrm p}$ both in the intervals with slow ($U_r<400\,\mathrm{km\,s}^{-1}$) and in the intervals with fast ($U_r>500\,\mathrm{km\,s}^{-1}$) solar wind. 
A potential reason for this behaviour lies in the fact that $\mathcal F_B$ does not directly depend on $U_r$, while $\mathcal F_{\mathrm p}$ changes significantly with $U_r$ (see Figures~\ref{am_speed} and \ref{am_speed_field}).
We attribute the natural variations in the angular-momentum flux to changes in the solar-wind source regions \citep{schwenn06,tindale17}, deflections during the expansion  \citep{egidi69,siscoe69}, or large-scale fluctuations as part of the turbulent spectrum \citep{tu95,bruno13}. Furthermore, the complex field and flow geometries in the low corona and latitudinal variations of the source regions create variations in $\mathcal F_{\mathrm{tot}}$  \citep{finley17,reville17,boe20,finley20}. 

Figure~\ref{am_massflux} confirms that the scatter of data points around $\mathcal F_{\mathrm p}=0$ is more symmetric in the tenuous, low-$\mathcal G_{\mathrm p}$ wind than in the denser, high-$\mathcal G_{\mathrm p}$ wind. Assuming that all wind is ejected from the Sun with positive $\mathcal F_{\mathrm p}$ due to the Sun's sense of rotation\footnote{Realistically, the possibility exists that some wind streams are initiated with  $U_{\phi}<0$ at the Sun.}, negative $\mathcal F_{\mathrm p}$ can only be created by local stream interactions, which in turn must increase $\mathcal F_{\mathrm p}$ of neighbouring plasma in order to fulfil angular-momentum conservation. Our observation of a more symmetric distribution around $\mathcal F_{\mathrm p}=0$ at low $\mathcal G_{\mathrm p}$ is then consistent with a scenario in which these local stream interactions occur more effectively in low-$\mathcal G_{\mathrm p}$ wind.  In this scenario, the mixing of $\mathcal F_{\mathrm p}$ in low-$\mathcal G_{\mathrm p}$ wind leaves the high-$\mathcal G_{\mathrm p}$ wind as the main carrier of net angular-momentum flux. Figure~\ref{am_massflux} then suggests that most of the net angular-momentum flux in the solar wind is carried by dense slow wind with $\mathcal F_{\mathrm p}>0$ in the intermediate-to-high-$\mathcal G_{\mathrm p}$ range.

Our spectral analysis of $\mathcal F_{\mathrm p}$ and $\mathcal F_B$ in Section~\ref{sect_fourier} quantifies the variations of the angular-momentum flux depending on frequency $\nu$.
The averaging time (60 minutes) for our analysis of the angular-momentum flux outside of Section~\ref{sect_fourier} corresponds to $\nu=2.78\times 10^{-4}\,\mathrm{Hz}$, which lies well inside the accessible frequency range of the high-$\nu$ regime. Figure~\ref{fourier} does not exhibit a clear peak at the frequency associated with the Sun's equatorial rotation frequency, $\nu=2\pi \Omega_{\odot}\approx 4.7\times 10^{-7} \,\mathrm{Hz}$. The lack of such a feature suggests that the variability of the angular-momentum flux induced by repeated passes of source regions due to the Sun's rotation is negligible in our dataset.
However, a spectral analysis of variations on these long timescales is not fully reliable at this stage of the mission due to the short overall duration of the dataset and its patchiness. The smallest gaps in our dataset are of order a few minutes, while the longest gap is 18 days long. We note in this context that large-scale interplanetary structures such as co-rotating interaction regions, which could be responsible for variations on these timescales, typically decay after a few revolutions.
The breakpoint at $\nu\sim 10^{-5}\,\mathrm{Hz}$ corresponds to a timescale of $\sim 28\,\mathrm h$. Assuming that this variation is frozen into the solar-wind flow, this timescale corresponds to a convected length scale of $\sim 0.27\,\mathrm{au}$ according to Taylor's hypothesis \citep{taylor38} based on an average $U_r\approx 400\,\mathrm{km\,s}^{-1}$.  This scale is greater than the typical correlation length in solar-wind turbulence in the inner heliosphere  \citep{matthaeus82,bruno86,bruno13,bourouaine20}. 
We note, however, that the uncertainty in the visual determination of the breakpoint frequency permits values between $\sim 6\times 10^{-6}\,\mathrm{Hz}$ and $\sim 4\times 10^{-5}\,\mathrm{Hz}$ due to the noise in our Fourier spectrum. Length scales within the corresponding range of breakpoint scales are still greater than the typical correlation length. In addition, the determination of the correlation length itself is to a certain degree method-dependent. Some estimates for the correlation length of velocity fluctuations provide the same order of magnitude as the length scale associated with our breakpoint in $\mathrm{PSD}(\mathcal F_{\mathrm p})$ \citep{podesta08}, supporting the link between fluctuations in angular-momentum flux with the turbulent fluctuations in the solar wind. It is worthwhile to perform a scale-dependent study of the angular-momentum flux in the future, especially by separating the variations in the low-$\nu$ regime and in the high-$\nu$ regime.

Sporadic events, such as the compression on 2020-09-06, contribute to the natural variations of the angular-momentum flux. We do not remove events like these from our statistical analysis, although they introduce a bias as seen in Figures~\ref{am_timeseries} and~\ref{hist_angmom_part}. The on-average greater variations in slow solar wind are reflected by the greater variation of the proton contribution $\mathcal F_{\mathrm p}$ to the angular-momentum flux at small $U_r$ in Figure~\ref{am_speed}.  The mass-flux dependence of $\mathcal F_{\mathrm p}$ during the time interval associated with the compression region in Figure~\ref{am_massflux} reveals a way to separate dense intermittent structures for future analyses. Using this method, it will be interesting to investigate the contribution of co-rotating interaction regions, coronal mass ejections, and other transient mass-flux enhancements to the Sun's long-term angular-momentum loss.  
In fact, the amount of angular momentum carried by coronal mass ejections is still not well known. Since angular momentum is conserved, however, local angular-momentum enhancements by the deflection of background solar wind must be balanced via flows with opposite angular momentum elsewhere in space. Figure~\ref{am_massflux} shows that the compression region in our dataset carries approximately the same specific angular momentum per proton, $rU_{\phi}$, as the regular slow wind, yet with a higher density. This observation suggests that the compression region represents ``scooped-up'' solar-wind mass, which has not undergone a significant alteration in terms of the particle's angular momentum. As shown in Figure~\ref{am_speed_field}, the value of $\mathcal F_B$ in the compression region is comparable to the average value of $\mathcal F_B$ across our dataset. This observation supports our interpretation of the compression region as a density increase of otherwise unaltered background solar wind. Composition measurements of solar-wind transients have the potential to confirm this interpretation in the future.

Our measurements neglect the contribution of $\alpha$-particles to the angular-momentum flux. Given that their contribution to the local momentum flux is of order 20\% (even greater near the Sun), $\alpha$-particles can make a significant contribution to the total angular-momentum flux \citep{pizzo83,marsch84,verscharen15,finley21}. 
Both \citet{finley21} and \citet{liu21} suggest a reconstruction of the $\alpha$-particle component based on the field-alignment of the differential flow between protons and $\alpha$-particles. It is still an open question, however, whether the $\alpha$-particles are in general a net source of positive or negative angular-momentum flux. In addition, the proton beam component can carry a significant angular momentum flux \citep{finley21}, which we directly subsume through our using of total proton moments for $\rho$ and $\vec U$.
Our analysis also neglects stresses due to pressure anisotropies in the particle populations \citep{hundhausen70a,marsch82,verscharen19}. This contribution to the angular-momentum flux is not significant for most of the solar-wind plasma though.

At this early stage of the mission, it is difficult to quantify the measurement uncertainties of the instruments accurately. These estimates will become more reliable as the mission progresses. At  $U_r\lesssim 300\,\mathrm{km\,s}^{-1}$, the sensitivity of SWA/PAS decreases, so that the uncertainty in the proton moments increases. Therefore, we urge caution regarding the interpretation of the proton data (especially, the proton density) in this velocity range, which includes the compression region. 
Solar Orbiter's Radio and Plasma Waves \citep[RPW,][]{maksimovic20}  instrument provides an estimate of the local electron density based on the probe-to-spacecraft potential and quasi-thermal noise measurements \citep{khotyaintsev21} which is independent of the SWA/PAS density measurement. We find that, especially during intervals with low $U_r$, SWA/PAS provides a greater density value than RPW. In Appendix~\ref{app}, we repeat part of our analysis using the independent RPW density estimate. Although we find a reduction in the peaks of $\mathcal F_{\mathrm p}$ when using the RPW density, our conclusions hold regardless of the used density measurement.  RPW is also able to provide an independent measurement of $|\vec U|$ from the bias DC electric field combined with the measurement of $|\vec B|$ from MAG.  A careful cross-calibration between SWA/PAS and RPW both in terms of density and bulk-speed measurements will improve further studies in the future \citep{owen20,maksimovic20,walsh20}.

Early results from Parker Solar Probe report an increased azimuthal flow of the protons compared to the expectation of the \citet{weber67} model at heliocentric distances up to about 0.23\,au \citep{kasper19}.  Possible explanations for this observed ``angular-momentum paradox''  \citep{reville20} include the partitioning of angular-momentum flux between the different particle species, non-axisymmetric flows and pressure gradients, and pressure anisotropies. In our data, however, we do not find evidence for such a persistent positive azimuthal flow, albeit our data were recorded at distances greater than 0.59\,au and show a different distribution of solar-wind speeds compared to the study by \citet{kasper19}.  In the future, when Solar Orbiter's perihelion distance is reduced, it will be important to monitor the azimuthal flow of the protons more closely to compare with Parker Solar Probe's findings of super-rotational flows in the near-Sun solar wind.

\section{Concluding remarks}

A reliable quantification of the Sun's global angular-momentum loss requires measurements of the angular-momentum flux over long time intervals. In addition, we require distinctive measurements of typical equatorial and polar solar wind \citep{mccomas00,verscharen21} to complete the understanding of the global angular-momentum loss. 
At this early point in the mission, we cannot confidently ascertain our measurement interval as a representative sample of the solar wind's angular-momentum flux. Further long-term studies will become available during Solar Orbiter's mission lifetime. In later stages of the mission, the spacecraft will leave the plane of the ecliptic making observations of the angular-momentum flux of polar solar wind feasible.  Nevertheless, our study already confirms the potential for future detailed studies of the large-scale properties of the solar wind with the data from Solar Orbiter.


%

%

\begin{acknowledgements}
Solar Orbiter is a space mission of international collaboration between ESA and NASA, operated by ESA.
The Solar Orbiter Solar Wind Analyser (SWA) PAS was designed, created, and is operated under funding provided in numerous contracts from the UK Space Agency (UKSA), the UK Science and Technology Facilities Council (STFC), the Centre National d'\'Etudes Spatiales (CNES, France), the Centre National de la Recherche Scientifique (CNRS, France), and the Czech contribution to the ESA PRODEX programme. In particular, operations at UCL/MSSL are currently funded under STFC grant ST/T001356/1. 
The Solar Orbiter Magnetometer was funded by the UK Space Agency (grant ST/T001062/1).
The RPW instrument has been designed and funded by CNES, CNRS, the Paris Observatory, the Swedish National Space Agency, ESA-PRODEX, and all the participating institutes.
D.V.~is supported by STFC Ernest Rutherford Fellowship ST/P003826/1. D.V.~and D.S.~are supported by STFC Consolidated Grant ST/S000240/1.  A.J.F.~ is supported by the ERC Synergy grant ``Whole Sun'', \#810218. 
T.H.~is supported by STFC grant ST/S000364/1. Y.~V.~K.~ is supported by the Swedish National Space Agency grant 20/136. 
We appreciate helpful discussions at the ISSI Team ``Exploring The Solar Wind In Regions Closer Than Ever Observed Before''.  We appreciate helpful discussions with Silvia Perri, Christian M\"ostl, and the members of the Solar Orbiter in-situ science working group ``CMEs, CIRs, HCS and large-scale structures''.
\end{acknowledgements}

\begin{appendix}
\section{The angular-momentum and mass flux based on the RPW density}\label{app}

\begin{figure*}
 \centering
\includegraphics[width=\textwidth]{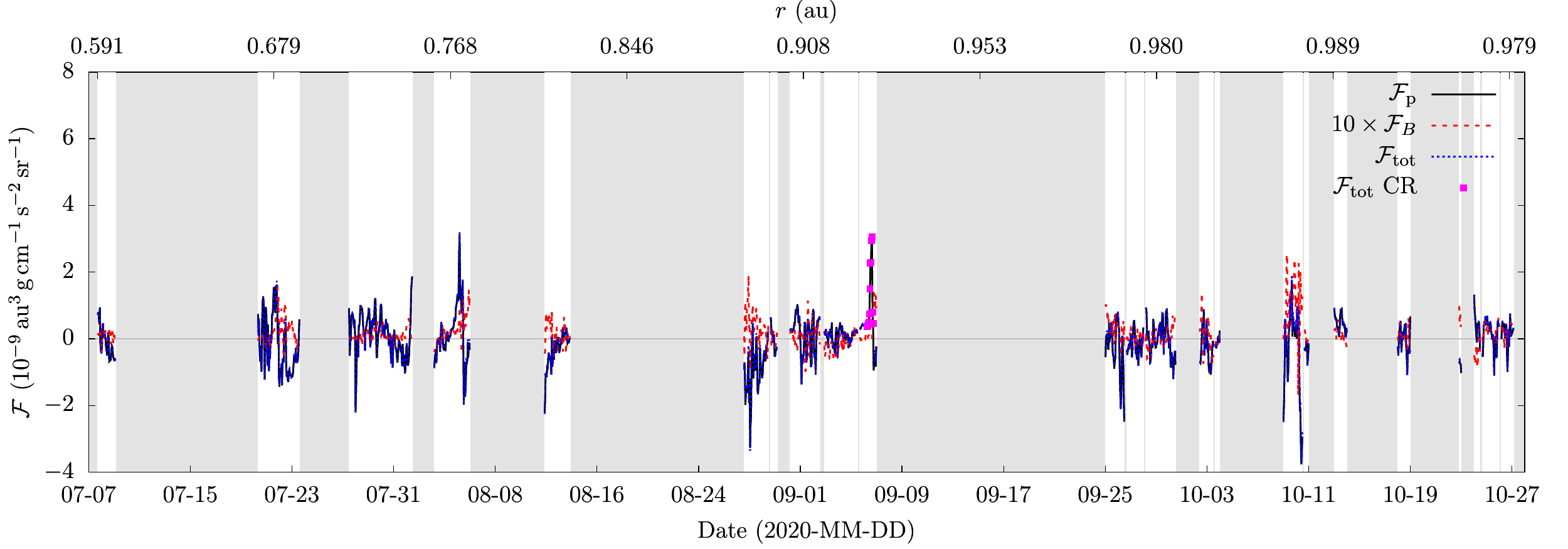}
\caption{Same as Figure~\ref{am_timeseries}, but using $\rho$ from RPW in the calculation of $\mathcal F_{\mathrm p}$ and $\mathcal F_{\mathrm{tot}}$. The grey-shaded areas indicate times for which our merged SWA/PAS-MAG-RPW data product is unavailable or the data flags for either dataset indicate poor data quality.}
   \label{am_timeseries_RPW}
 \end{figure*}

RPW measures the local total electron density $n_{\mathrm e}$ based on a combination of both the peak-tracking of the plasma frequency and the spacecraft potential \citep{khotyaintsev21}. Due to quasi-neutrality, $n_{\mathrm e}$ serves as an independent measure of the total charge-weighted ion density. For comparison with our SWA/PAS-MAG measurements, we create a new dataset using $\vec U$ from SWA/PAS, $\vec B$ from MAG, and $\rho=m_{\mathrm p}n_{\mathrm e}$ from RPW, where $m_{\mathrm p}$ is the proton mass. We apply the same selection, averaging, and analysis methods as in the main part of this work. This leaves us with 913 data points for our combined SWA/PAS-MAG-RPW dataset in total.

We show the timeseries of our combined SWA/PAS-MAG-RPW dataset in Figure~\ref{am_timeseries_RPW}. Qualitatively, the timeseries agrees with Figure~\ref{am_timeseries}. However, the peaks with $\mathcal F_{\mathrm p}>0$ are less pronounced in the SWA/PAS-MAG-RPW dataset than in the SWA/PAS-MAG dataset. These peaks, including our compression region on 2020-09-06, correspond to intervals of slow and dense solar wind, for which the SWA/PAS and RPW densities diverge most. 

We present the statistical properties of our combined SWA/PAS-MAG-RPW dataset in Table~\ref{table_stats_RPW}. Due to the lower peaks in $\mathcal F_{\mathrm p}$, the signs of the mean values for $\mathcal F_{\mathrm p}$ and $\mathcal F_{\mathrm{tot}}$ across our full dataset are now different from the signs of the SWA/PAS-MAG dataset shown in Table~\ref{table_stats}. The maxima of $\mathcal F_{\mathrm p}$ and $\mathcal F_{\mathrm{tot}}$ in Table~\ref{table_stats_RPW} are lower than those given in Table~\ref{table_stats}, reflecting the lower peaks seen in Figure~\ref{am_timeseries_RPW}. All other statistical markers in Table~\ref{table_stats_RPW} agree with our results shown in Table~\ref{table_stats}.

 \begin{table*}
 \caption{Same as Table~\ref{table_stats}, but using $\rho$ from RPW in the calculation of $\mathcal F_{\mathrm p}$ and $\mathcal F_{\mathrm{tot}}$.   }
 \label{table_stats_RPW}
 \centering
 \begin{tabular}{lccc}
 \hline \hline
  & mean & min & max \\
  &$(10^{-11}\,\mathrm{au}^3\,\mathrm g\,\mathrm{cm}^{-1}\,\mathrm s^{-2}\,\mathrm{sr}^{-1})$ & $(10^{-9}\,\mathrm{au}^3\,\mathrm g\,\mathrm{cm}^{-1}\,\mathrm s^{-2}\,\mathrm{sr}^{-1})$&$(10^{-9}\,\mathrm{au}^3\,\mathrm g\,\mathrm{cm}^{-1}\,\mathrm s^{-2}\,\mathrm{sr}^{-1})$ \\
 \hline
 all speeds & & & \\
 $\mathcal F_{\mathrm{p}}$ & $-5.33\pm 72.20$ & $-3.76$ & $3.12$  \\
$  \mathcal F_{B}$ & $1.67\pm4.24$ & $-0.169$ & $0.254$ \\
$ \mathcal F_{\mathrm{tot}}$ & $-3.66\pm72.20$ & $-3.69$ & $3.21$ \\
 \hline
 $U_r<400\,\mathrm{km\,s}^{-1}$ & & & \\
 $\mathcal F_{\mathrm{p}}$ & $10.5\pm 70.5$ & $-2.49$ & $3.12$  \\
$  \mathcal F_{B}$ & $0.735\pm5.840$ & $-0.230$ & $0.254$ \\
$ \mathcal F_{\mathrm{tot}}$ & $11.2\pm70.3$ & $-2.49$ & $3.21$ \\
 \hline
  $U_r>500\,\mathrm{km\,s}^{-1}$ &  & & \\
 $\mathcal F_{\mathrm{p}}$ & $-67.2\pm 106.5$ & $-5.43$ & $0.753$  \\
$  \mathcal F_{B}$ & $0.605\pm5.464$ & $-0.174$ & $0.101$ \\
$ \mathcal F_{\mathrm{tot}}$ & $-66.6\pm108.6$ & $-5.57$ & $0.836$ \\
 \hline
 \end{tabular}
 \end{table*}

\begin{figure}
 \centering
\includegraphics[width=\columnwidth]{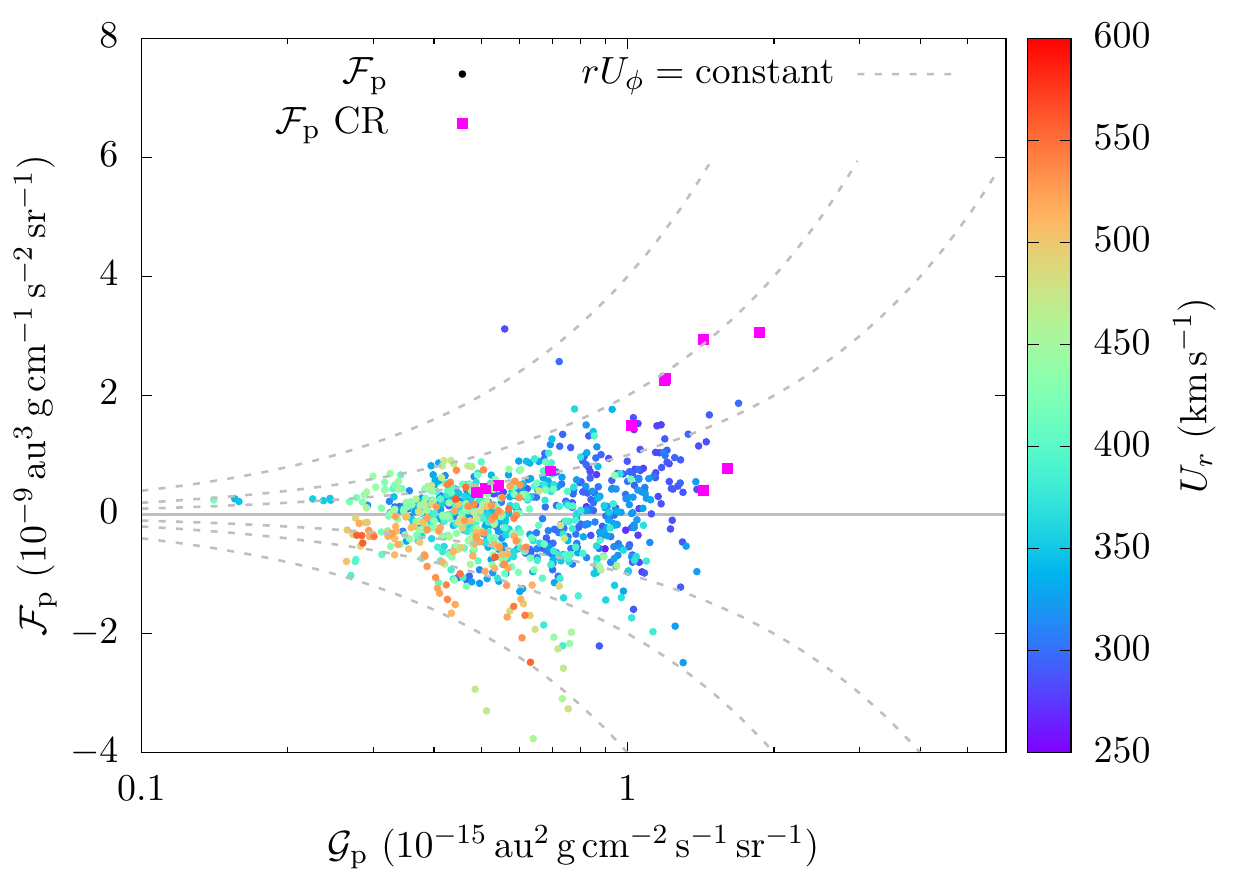}
\caption{Same as Figure~\ref{am_massflux}, but using $\rho$ from RPW  in the calculation of $\mathcal F_{\mathrm p}$ and $\mathcal G_{\mathrm p}$. 
}
   \label{am_massflux_RPW}
 \end{figure}

We show the dependence of $\mathcal F_{\mathrm p}$ on $\mathcal G_{\mathrm p}$ for our combined SWA/PAS-MAG-RPW dataset in Figure~\ref{am_massflux_RPW}. The comparison between Figures~\ref{am_massflux} and \ref{am_massflux_RPW} shows that most high-$\mathcal G_{\mathrm p}$ measurements, including the compression region, shift towards smaller values of $\mathcal G_{\mathrm p}$ when using the RPW density estimate. In addition, the compression region shifts towards smaller values of $\mathcal F_{\mathrm p}$. Notwithstanding these differences, our conclusions drawn based on Figure~\ref{am_massflux} are still valid.

\end{appendix}

\bibliographystyle{aa} 
\bibliography{solo_am} 

\end{document}